\documentclass[twocolumn,preprintnumbers,amsmath,amssymb,prl]{revtex4}

\usepackage{graphicx}
\usepackage{color}

\begin{document}

\title{Wideband tunable, graphene-mode locked, ultrafast laser}
\author{Z. Sun$^1$, D. Popa$^1$, T. Hasan$^1$, F. Torrisi$^1$, F. Wang$^1$, E. J. R. Kelleher$^2$, J. C. Travers$^2$, A. C. Ferrari$^1$}
\email{acf26@eng.cam.ac.uk}
\affiliation{$^1$Department of Engineering, University of Cambridge,Cambridge CB3 0FA, UK\\
$^2$Physics Department, Imperial College, London SW7 2AZ, UK}


\begin{abstract}
We report a tunable ultrafast fiber laser mode-locked with a graphene saturable absorber. The linear dispersions of the Dirac electrons in graphene enable wideband tunability. We get$\sim$1ps pulses, tunable between 1525 and 1559nm, demonstrating graphene as a broadband saturable absorber.
\end{abstract}

\maketitle

Ultrafast passively mode-locked fiber lasers with spectral tuning capability have widespread applications in biomedical research, spectroscopy and telecommunications\cite{Letokhov_nature,Okhotnikov_njp,Okhotnikov_ol_2003}, due to their simplicity, compactness and efficient heat dissipation\cite{Okhotnikov_njp,Okhotnikov_ol_2003,Ober_OL_1995,Tamura_el_1995,Tamura_ptl_94_2}. Currently, the dominant technology is based on semiconductor saturable absorber mirrors (SESAMs)\cite{Okhotnikov_njp,Okhotnikov_ol_2003,Ober_OL_1995}. However, these have a narrow tuning range, and require complex fabrication and packaging\cite{Keller_nature_03,Okhotnikov_njp}. A simple, cost-effective alternative is to use nanotubes (CNTs)\cite{Wang_nn_2008,Set_stqe_03,Rozhin_pssb_2006,Della_apl_06,Scardaci_pe_2007,Scardaci_am_2008,Sun_apl_2008,Sun_apl_2009,Kelleher_apl_09,Kelleher_ol_09,Kieu_oe_08,Solodyankin_ol_08,Nicholson_oe_07,Song_apl_08,Shohda_oe_09,Schmidt_oe_09,Fong_ol_07,Hasan_am_2009}, In CNTs the diameter controls the bandgap, thus defining the operating wavelength. Broadband tunability is possible using CNTs with a wide diameter distribution\cite{Wang_nn_2008,Senoo_oe_09,Kivisto_oe_09}. However, when operating at a particular wavelength, CNTs not in resonance are not used and give insertion losses.

After the first demonstration of a graphene-based mode-locker\cite{Hasan_am_2009}, a variety of lasers were reported exploiting graphene saturable absorbers for ultrafast pulse generation at 1 and 1.5$\mu$m\cite{Sun_acsnano_10,Zhang_oe_09,Bao_afm_09,Zhang_apl_09,Song_apl_09,Tan_apl_10,Bao_afm_10,Zhang_apl_10_tun}. Ref.\onlinecite{Sun_acsnano_10} explained the fundamentals of the photo-excited carrier dynamics, which leads to Pauli-blocking and, thus, saturable absorption, with good agreement between theory and experiments. The linear dispersion of the Dirac electrons in graphene provides an ideal solution for wideband ultrafast pulse generation\cite{Sun_acsnano_10}. Wavelength tuning in graphene-based lasers has been achieved by exploiting fiber birefringence\cite{Bao_afm_10,Zhang_apl_10_tun}. However, fiber birefringence is sensitive to temperature fluctuations and other environmental instabilities\cite{Agrawal_book_app}, making this approach non ideal for long-term-stable systems; a key requirement for mode-locked lasers used in practical applications.

Here, we demonstrate an ultrafast tunable fiber laser mode-locked by a graphene-based saturable absorber, with stable mode-locking over 34~nm, insensitive to environmental perturbations. The tuning range is limited only by the tunable filter, with a wider range obtainable with a broader filter. The output pulse duration is $\sim$1~ps.

The saturable absorber is prepared as described in Ref.\onlinecite{Sun_acsnano_10}. Graphene flakes are exfoliated by mild ultrasonication with sodium deoxycholate surfactant. The dispersion is enriched with single (SLG) and few layer graphene (FLG), and mixed with polyvinyl alchohol (PVA; Wako chemicals). Slow water evaporation in a dessicator produces free-standing, 50~$\mu$m thick graphene-PVA (GPVA) composites\cite{Hasan_am_2009,Sun_acsnano_10}. Absorption and Raman spectroscopy are used for their characterization\cite{Ferrari_prl_06,Sun_acsnano_10}. Apart from the characteristic $\pi$ absorption in the UV region\cite{eberlein_prb_2008} and the host matrix PVA, the GPVA films are featureless over a 1500~nm range (from $\sim$500 to $\sim$2000nm). Raman spectroscopy indicates the layers in the FLG flakes behave like electronically de-coupled SLG, retaining the linear dispersion of Dirac fermions\cite{Sun_acsnano_10,Ferrari_prl_06,latil_prb_2007}. Compared to other graphene saturable absorber fabrication strategies\cite{Zhang_oe_09,Bao_afm_09,Zhang_apl_09,Song_apl_09,Tan_apl_10,Bao_afm_10,Zhang_apl_10_tun}, our approach is easily scalable, and allows integration into various photonic systems\cite{Hasan_am_2009,Sun_acsnano_10}.

\begin{figure}
\centerline{\includegraphics[width=75mm]{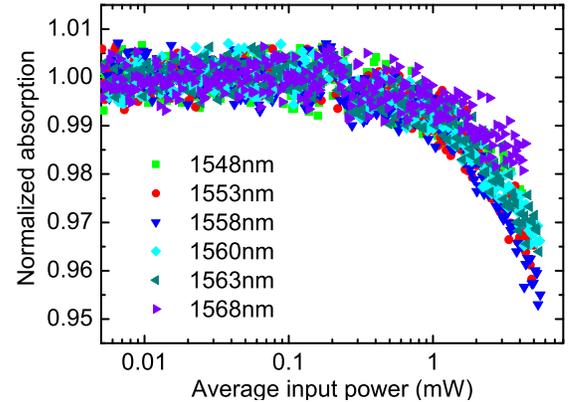}} \caption{\label{absorption}Power-dependent absorption at six wavelengths. Input repetition rate$\sim$38MHz, pulse duration$\sim$580fs}
\end{figure}

Power-dependent absorption at six wavelengths is measured using an all-fiber based setup described in Ref.\onlinecite{Hasan_am_2009}, Fig.\ref{absorption}, by sandwiching the GPVA film between two fiber connectors. The absorption decreases by$\sim$4.5\% due to saturation when the incident average power is increased to 5.35mW (266MW/cm$^{2}$ power density) at 1558nm, independent of wavelength. The nonlinear operation in terms of modulation depth and non-saturable absorption of our graphene saturable absorber is comparable to that of the CNT-based devices reported in Refs.\onlinecite{Wang_nn_2008,Scardaci_am_2008,Sun_apl_2008,Sun_apl_2009}. Note that the data in Fig.1 are just limited by our pump wavelengths availability. Saturable absorption is expected over a much wider spectral range due to the linear dispersion of the graphene electrons\cite{Hasan_am_2009,Sun_acsnano_10}.
\begin{figure}
\centerline{\includegraphics[width=75mm]{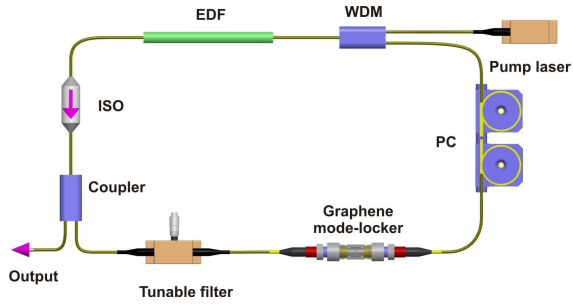}} \caption{\label{Setup} Tunable laser setup}
\end{figure}
\begin{figure}
\centerline{\includegraphics[width=75mm]{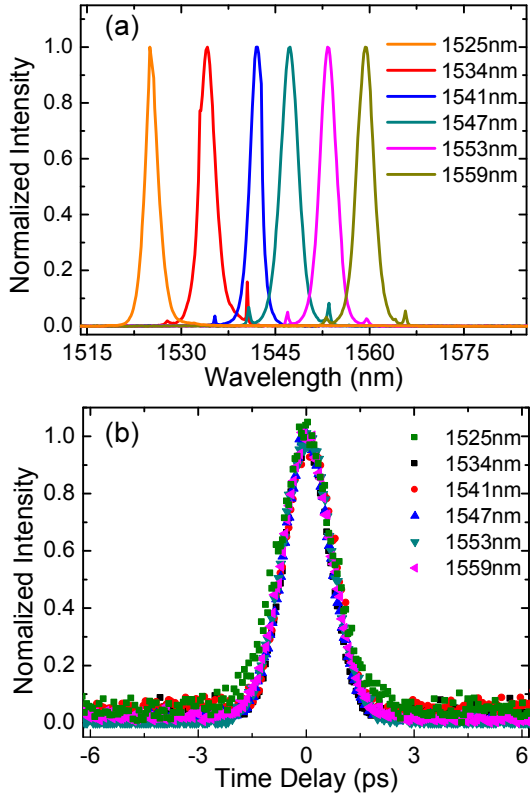}}\caption{\label{spectrum_AC}(a)Output spectra. (b)Autocorrelation traces at different wavelengths.}
\end{figure}
\begin{figure}
\centerline{\includegraphics[width=75mm]{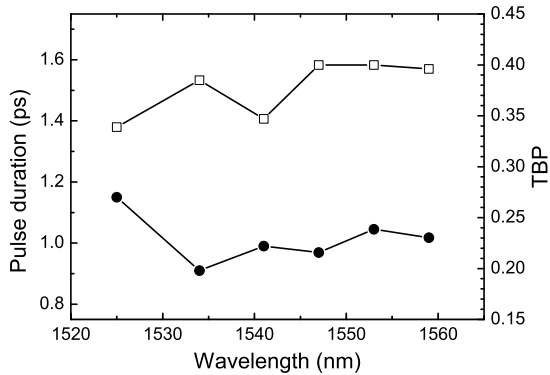}}
 \caption{\label{TBP}(Circles) Output pulse duration; (squares) time-bandwidth product.}
\end{figure}
\begin{figure}
\centerline{\includegraphics[width=80mm]{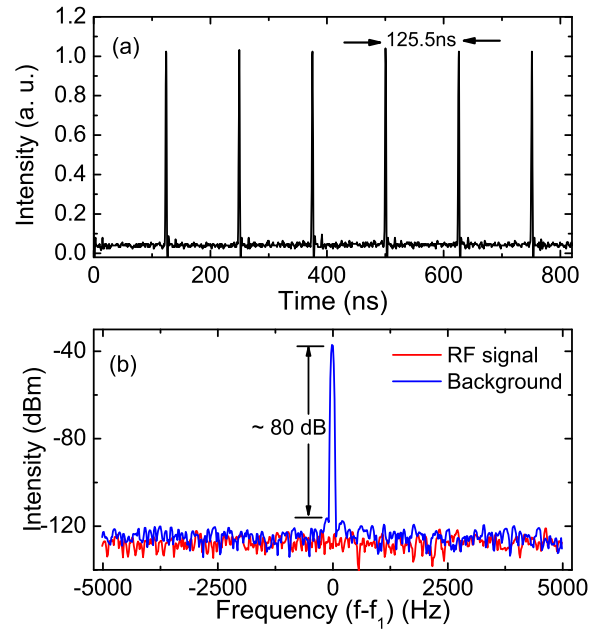}} \caption{\label{RF}(a)Output pulse train;(b)RF spectrum, measured around the fundamental repetition rate $f_{1}\sim$8MHz over 10kHz with 10Hz resolution.}
\end{figure}

The packaged GPVA is then integrated to form a tunable ultrafast laser, Fig. \ref{Setup}. The cavity consists of an Erbium doped fiber (EDF), followed by an optical isolator (ISO), a fused fiber coupler, a tunable filter with a 12.8~nm bandwidth, a polarization controller (PC), and a fused fiber wavelength division multiplexer (WDM). A 1.2m EDF (Fibercore) is backward pumped by a 976nm laser diode (Bookham LC96V74-20R) through the WDM. Unidirectional operation is imposed by the ISO. The PC is used to adjust polarization for mode-locking optimization. The 20\% port of the 20/80 coupler feeds the pulses back into the cavity. The total cavity length is$\sim$26.3m. The lasing wavelength is selected by the in-line tunable filter. The band-pass filter angle is controlled by a micrometer screw, providing continuous tunability from 1530 to 1555nm. This ensures stable mode-locking, independent of temperature and environmental fluctuations, unlike the wavelength selection strategy based on tuning fiber birefringence with a polarization controller\cite{Bao_afm_10,Zhang_apl_10_tun}. The pump and average output power are monitored by a photodiode, while the pulse duration and spectrum are recorded using a second harmonic generation (SHG) intensity autocorrelator (APE Pulsecheck 50) and an optical spectrum analyzer (Anritsu MS9710B), respectively.

Continuous wave operation starts at$\sim$6mW pump power, giving an output power$\sim$35$\mu$W. Single-pulse mode-locking is observed at$\sim$20mW pump power. The output wavelength is tunable from 1525 to 1559nm, Fig.\ref{spectrum_AC}(a), limited by the tunable filter used in our experiment, not by the graphene-saturable absorber. The full width at half maximum (FWHM) spectral bandwidth at a representative output wavelength (1553nm) is$\sim$3nm. Typical soliton sidebands are observed, due to periodic intracavity perturbations\cite{Dennis_jqe_1994}. The autocorrelation traces are shown in Fig. \ref{spectrum_AC}(b). Assuming a sech$^2$ pulse shape, the de-convolved pulse duration is$\sim$1ps at 1553nm. Pulse durations and time-bandwidth products (TBP) at different output wavelengths are shown in Fig.\ref{TBP}. The TBP at 1553nm is$\sim$0.4. The deviation from 0.315, expected for transform-limited sech$^{2}$ pulses, indicates minor chirping\cite{Agrawal_book_app}. The nominal tuning range of our filter is 1530-1555nm, but it can also work from 1525 to 1530nm and from 1555 to 1559nm, at the expense of increased insertion losses. Thus, the achieved shortest and longest wavelengthes using this filter are 1525 and 1559nm. The pulse at 1525nm is slightly longer,$\sim$1.15ps, possibly due to the higher losses. The average output power is$\sim$1mW, with 125pJ pulse energy.

The stability is characterized from radio-frequency (RF) measurements of the output intensity\cite{Von1986}. The repetition rate is$\sim$8MHz, corresponding to$\sim$125.5ns round-trip time, Fig.\ref{RF}(a). Fig.~\ref{RF}(b) shows the RF spectrum of the fundamental harmonic frequency. The peak to pedestal extinction is$\sim$80dB, indicating low amplitude noise fluctuations\cite{Von1986}. Compared to previous tunable, graphene mode-locked lasers\cite{Bao_afm_10,Zhang_apl_10_tun}, we achieve significantly lower fluctuations and shorter pulse durations, showcasing the wideband operation capability of graphene and its potential as a wideband mode-locker.

We acknowledge funding from a Royal Society Brian Mercer Award for Innovation, the ERC grant NANOPOTS, EPSRC grants EP/GO30480/1, EP/G042357/1, Kings College and Imperial College.

\end{document}